\documentclass[twocolumn]{aastex61}
\pdfoutput=1 
\usepackage{amsmath}

\shorttitle{A More Informative Map}
\shortauthors{Rauscher, Suri, \& Cowan}

\begin{document}

\title{A More Informative Map: Inverting Thermal Orbital Phase \\
	and Eclipse Lightcurves of Exoplanets}

\email{erausche@umich.edu}

\author[0000-0003-3963-9672]{Rauscher, Emily}
\affil{Department of Astronomy, University of Michigan, 1085
  S. University Ave., Ann Arbor, MI 48109, USA}
\author{Suri, Veenu}
\affil{Department of Astronomy, University of Michigan, 1085
  S. University Ave., Ann Arbor, MI 48109, USA}
\author[0000-0001-6129-5699]{Cowan, Nicolas B.}
\affiliation{Department of Earth \& Planetary Sciences, McGill University, 3450 rue University, Montreal, QC, H3A 0E8, CAN \\ Department of Physics, McGill University, 3600 rue University, Montreal, QC, H3A 2T8, CAN.\\
McGill Space Institute, 3550 rue University, Montreal, QC, H3A 2A7, CAN.\\
Institut de recherche sur les exoplan\`etes, Universit\'e de Montr\'eal, C.P. 6128, Succ.\ Centre-ville, Montr\'eal QC H3C 3J7, CAN}
  
\begin{abstract}

Only one exoplanet has so far been mapped in both longitude and latitude, but the James Webb Space Telescope should provide mapping-quality data for dozens of exoplanets.  The thermal phase mapping problem has previously been solved analytically, with orthogonal maps---spherical harmonics---yielding orthogonal lightcurves---sinusoids.  The eclipse mapping problem, let alone combined phase+eclipse mapping, does not lend itself to such a neat solution. Previous efforts have either adopted spherical harmonics, or various ad hoc map parameterizations, none of which produce orthogonal lightcurves.   We use principal component analysis to construct orthogonal ``eigencurves," which we then use to fit published 8 micron observations of the hot Jupiter HD~189733b.  This approach has a few advantages over previously used techniques: 1) the lightcurves can be pre-computed, accelerating the fitting process, 2) the eigencurves are orthogonal to each other, reducing parameter correlations, and 3) the eigencurves are model-independent and are ranked in order of sensitivity.  One notable result of our analysis is that eclipse-only mapping of HD~189733b is far more sensitive to the central concentration of dayside flux than to the eastward offset of that hotspot.  Mapping can, in principle, suffer from degeneracies between spatial patterns and orbital parameters.  Previous mapping efforts using these data have either assumed a circular orbit and precise inclination, or have been pessimistic about the prospects of eclipse mapping in the face of uncertain orbital parameters. We show that for HD~189733b the combined photometry and radial velocity are sufficiently precise to retire this concern.   Lastly, we present the first map of brightness temperature, and we quantify the amplitude and longitude offset of the dayside hotspot. 

\end{abstract}


\section{Introduction}

There is so far only one exoplanet whose brightness structure we have been able to map in both latitude and longitude \citep{deWit2012,Majeau2012}, by stacking together multiple Spitzer Space Telescope observations of one of the brightest hot Jupiters. With the imminent launch of the James Webb Space Telescope, there soon should be data of sufficient quality to easily map bright transiting exoplanets \citep{Rauscher2007b,Schlawin2018}. These maps will be produced by converting measurements of flux as a function of time to flux as a function of spatial location, through the physical processes of the planet's rotation bringing different regions into view and the stellar limb sequentially hiding/revealing different slices of the planet's day side as it enters/exits secondary eclipse, the period of time it spends hidden behind the star. We call these types of measurements orbital phase curves and eclipse mapping observations---a review of the mapping methods is presented in \citet{Cowan_Fujii2017}.

In practice, the process of inverting flux curve observations into spatial flux maps is complex, with results that can depend sensitively on the assumed map structure \citep{deWit2012,Majeau2012,Louden2018} and have spatial patterns degenerate with uncertainties in the planet's orbital parameters \citep{Williams2006,deWit2012}. 
Given the subtle nuances inherent in the mapping process, at the dawn of the new JWST era of exoplanet characterization, here we present an analysis method that can maximize the information reliably retrieved from flux curve observations, taking into account uncertainties in the orbital system parameters in a fully consistent way. 
Here we focus on the case of emitted flux from a planet that is assumed to be tidally locked into synchronous rotation (so that it is a simple conversion from orbital phase to observed longitude on the planet), but our approach could be expanded to include reflected light and non-synchronous rotation in future work. We advocate the use of this technique on upcoming JWST data sets of thermal emission from hot Jupiters and other close-in exoplanets.

After reviewing previous literature on mapping exoplanets (Section \ref{sec:history}), we describe our method. We begin by determining the optimized flux curves to use in fitting data, as described in Section~\ref{sec:eigencurves}. We then discuss some of the features of these optimized flux curves and their corresponding maps in Section~\ref{sec:eigenmaps}. In Section~\ref{sec:orbiterrors} we demonstrate how uncertainties in orbital system parameters can be folded into this method, but that it is also possible to calculate ahead of time whether parameter uncertainties will induce significant uncertainties (or not) in retrieved maps. Finally, in Section~\ref{sec:data} we apply our technique to the only planet for which we currently have data of high enough precision to produce a map, HD~189733b, and compare our result to the maps previously presented in \citet{deWit2012,Majeau2012}. In 
Section~\ref{sec:summary} we summarize our method and discuss some of the nuances of creating maps.

\section{A Brief History of Exocartography} \label{sec:history}
\subsection{Thermal Phase Curves}
\cite{Harrington2006} and \cite{Cowan2007} reported multi-epoch \emph{Spitzer} phase curves of non-transiting hot Jupiters, allowing the authors to constrain the day/night contrast of those planets.  But it was the continuous phase monitoring of \cite{Knutson2007b} that enabled the first bona fide one-dimensional (longitudinal) map of an exoplanet: the authors used a 12-slice model constrained by regularization.  \cite{Cowan2008} presented the details of that approach, and developed the analytic formalism for a Fourier-based approach.  They found that for planets with edge-on orbits, there is a one-to-one correspondence between the Fourier modes of a planet's orbital phase variations and the longitudinal brightness map of the planet.  They further showed that the map-to-lightcurve transformation has a nullspace: odd harmonics should not be present in the thermal lightcurve of an edge-on planet.

\cite{Cowan2013a} analytically solved the general thermal phase curve problem and found that even transiting planets, which typically have orbital inclinations within a few degrees of edge-on, can exhibit odd harmonics if they have north-south asymmetric maps. \cite{Cowan2017} subsequently demonstrated that time-variable maps could also produce odd harmonics. No matter how you cut it, odd harmonics in the phase curve of a planet on a circular orbit indicate climatic features: spatially localized and/or time-varying atmospheric phenomena.  So far, these modes have only been reliably reported in the \emph{Kepler} lightcurves of the hot Jupiters Kepler-13Ab and HAT-P-7b \citep{Esteves2015} and may be due to slight eccentricity of those systems \citep{Penoyre2018}.  For the purposes of this paper, we presume that planet brightness variations are due to the inhomogeneous thermal emission of a spherical planet on a circular orbit, although we do account for the possibility of non-zero eccentricities (Section \ref{sec:orbiterrors}).

Researchers have continued to obtain, reduce, and analyze thermal phase curves of short-period planets \citep[for a review see][]{Parmentier2017}.  But with a few exceptions \citep{Knutson2009a,Cowan2012a,Dang2018}, map-making fell out of favor: observers would present phase curves detrended for instrument effects and theorists would contribute disk-integrated predictions from their general circulation models.  In hindsight, skipping the mapping step turns out to have been ill-advised: \cite{Keating2017} showed how the published phase curves of WASP-43b implied negative brightnesses at certain longitudes on the planet, presumably due to incomplete detrending of detector systematics \citep[indeed, subsequent reanalysis of the WASP-43b phase curves found different, physically allowed, solutions:][]{Louden2018,Mendoca2018}.  Although it was previously acknowledged that the disk-integrated brightness of a planet, i.e., its phase curve, must be non-negative, \cite{Keating2017} showed that even strictly positive phase curves can be unphysical if they imply negative brightness at certain longitudes.  In other words, one should always convert phase curves into maps, if only to ensure that the phase curve is physically allowed. Fortunately, it is easy to invert phase curves into planetary maps using either analytic deconvolution \citep{Cowan2008} or fast numerical methods  \citep[SPIDERMAN;][]{Louden2018}. 

Lastly, we note that the mathematics and science of exoplanet thermal phase curves has much in common with the rotational modulation of brown dwarfs \citep[see review by][]{Artigau2018}.

\subsection{Reflected Phase Curves}
While phase curves at longer wavelengths probe the thermal emission of a planet, optical wavelengths are often more sensitive to reflected light.  As such, optical phase curves can be used to constrain the albedo map of an exoplanet.  There is a long history, dating back to \cite{Russell1906}, of inverting time-variable brightness of an unresolved astronomical object  to infer its albedo markings.  Most modern reflected-light efforts were directed towards next-generation direct-imaging missions \citep[e.g.,][]{Oakley2009,Cowan2009,Fujii2012}, but they can readily be adapted to the simpler geometry of short-period planets, which are thought to have zero obliquity and to be synchronously rotating.

The contrast ratios tend to be more daunting than for thermal phase curves, but this science has been made possible by the Kepler mission \citep{Borucki2010}.  The first optical map of an exoplanet was reported by \cite{Demory2013} and this was converted into an albedo map by \cite{Cowan_Fujii2017}.   The edge-on reflected lightcurve problem was solved analytically by \cite{Cowan2013a}, and the general solution was developed by \cite{Haggard2018}.  \cite{Kawahara2010,Kawahara2011}, \cite{Fujii2012}, and \cite{Farr2018} have presented numerical approaches to inferring the albedo map of a planet based on its time-variable photometry.  These codes could be adapted to the simpler geometry of synchronously rotating planets.

It should be noted that the distinction between reflected light and thermal emission is slippery for highly-irradiated planets. With temperatures of 1000~K or more, there can be considerable thermal emission at optical wavelengths, and reflected starlight in the near infrared \citep{Schwartz2015}.  For example, \cite{Keating2017} found that up to half of the NIR dayside ``emission'' of WASP-43b was potentially reflected light \citep[see also][]{Louden2018}.  In the current manuscript, we only consider thermal emission, which greatly simplifies the problem and is a safe approximation for the 8~micron observations of the hot Jupiter HD~189733b that we analyze with our mapping method. 

\subsection{Eclipse Mapping}
\cite{Williams2006} were the first to note that the time of eclipse relative to transit is sensitive to the dayside brightness distribution of a planet.  Unfortunately, this signal is largely degenerate with the sky-plane component of the planet's eccentricity, $e\cos\omega$.  Nonetheless \cite{Agol2010} showed that the 30~second delay in the 8~micron eclipse of HD~189733b was consistent with the eastward offset of its dayside hotspot as inferred from phase curves, assuming the planet has zero eccentricity.  Or to put it differently, the combination of phase curve and time of eclipse puts an upper limit on the planet's $e\cos\omega$ \citep[Likewise, the \emph{duration} of eclipse relative to the duration of transit constrains $e\sin\omega$;][]{Winn2010}.  Insofar as different wavelengths of light originate from different depths in the atmosphere and hence exhibit different brightness maps, multi-wavelength eclipse data can provide constraints on the dayside brightness distribution of a planet, even in the absence of precise orbital constraints \citep{deWit2012,Dobbs-Dixon2015}.   

In addition to the exact time of eclipse, \cite{Rauscher2007b} showed that the \emph{morphology} of ingress and egress is sensitive to the 2D dayside map of the exoplanet.  This effect was subsequently reported for the hot Jupiter HD~189733b by \cite{Majeau2012} and \cite{deWit2012}. The lowest-order component of this signal---the duration of ingress and egress---is largely degenerate with the impact parameter of the planet at eclipse, which in turn is a function of the line-of-sight orbital eccentricity, $e\sin\omega$. \cite{Wong2014} performed the only other serious attempt to use the morphology of eclipse ingress and egress to map an exoplanet; they obtained only upper limits. Ironically, they were able to use the out-of-eclipse baseline---in effect a partial phase curve---to put useful constraints on the brightness variations of the planet.  This goes to show that the distinction between phase and eclipse observations is blurry \citep[e.g.,][]{Placek2017}. 

\subsection{Spectral Mapping}
\cite{Stevenson2014} reported the first phase-resolved spectra of an exoplanet, from which they retrieved the vertical temperature structure of the planet as a function of sub-observer longitude. Using the same data, the authors also produced longitudinal maps as a function of wavelength.  
High quality spectral eclipse data will likewise enable the construction of two-dimensional maps (longitude and latitude) as a function of wavelength. 
In particular, the large collecting area and spectral coverage of JWST is expected to produce high-quality spectral eclipse data for hot Jupiters \citep{Beichman2014,Cowan2015}.

If there were a one-to-one correspondence between wavelength and pressure, then one could convert spectrally-resolved maps into three-dimensional temperature maps of the planet. In reality, the vertical contribution function is complex and changes dramatically from one location to the next \citep{Dobbs-Dixon2017}. Thus, 3D mapping either requires a 3D spectral retrieval code, or performing 1D spectral retrieval at various locations of a multi-band map.  In this paper we tackle the mapping problem at a single wavelength, a prerequisite to the latter approach.

\section{Eigencurves} \label{sec:eigencurves}

One of the inherent tensions in the reconstruction of spatial brightness information on the planet from its emitted light curve is that a set of orthogonal basis maps on the planet (e.g., spherical harmonics) does not necessarily correspond to a set of orthogonal light curves. Using a set of non-orthogonal light curves to fit to a data set then necessarily scrambles the information available about the planet map. \cite{Majeau2012} and \cite{deWit2012} used spherical harmonics as their basis to fit the observations of HD~189733b and hence their lightcurves were not orthogonal. \citep[][also experimented with ad hoc models for which neither the maps nor the lightcurves are orthogonal]{deWit2012}. 

The method that we describe here starts with spherical harmonic maps to create light curves, but then uses a principal component analysis (PCA) of those light curves to calculate what we call ``eigencurves'', a set of light curves that are orthogonal.

We use the SPIDERMAN code presented in \citet{Louden2018} to create a set of light curves, each produced from a different spherical harmonic map. SPIDERMAN has several pre-defined planet map options, including spherical harmonics, and calculates the integrated light from the visible hemisphere of a planet as a function of time, for user-defined system parameters. This numerical integration is necessary \citep[instead of using analytic expressions from][]{Cowan2013a} because of the ingress and egress times of secondary eclipse, when the precise system geometry determines which region of the planet's disk is eclipsed by the stellar limb. Accurately capturing the secondary eclipse is important, since most latitudinal information comes from these times in the light curve.

We use the system parameters of HD~189733b (Table~\ref{tab:params}) to calculate these light curves. This choice allows us to compare to the data for this system in Section~\ref{sec:data}, but our method is generally applicable to any transiting system. We specifically discuss the impact of uncertainties in orbital parameters in Section~\ref{sec:orbiterrors} but for now assume precise values.

\begin{deluxetable}{lcc}
\tablecaption{HD 189733 system parameters}
\tablehead{\colhead{Parameter} & \colhead{Value} & \colhead{Uncertainty}} 
\startdata
Orbital period [days] &  2.21857567 & 0.00000015 \\
Semi-major axis, $a$ [AU] &  0.0313 & 0.0004 \\
Orbital inclination [degrees] &  85.710 & 0.024 \\
Eccentricity\tablenotemark{a} & 0 & 0.000094 \\
Planet-to-star radius ratio ($R_p/R_*$) &  0.155313 & 0.000188 \\
Scaled semi-major axis ($a/R_*$) &  8.863 & 0.020 \\
Stellar effective temperature [K] & 5052 & 16 \\
\enddata
\tablecomments{All values taken from \citet{Agol2010} except the semi-major axis \citep[from][]{Bouchy2005} and the stellar temperature \citep[from][]{Stassun2017}. The impact of the uncertainties in these values on our retrieved map is explored in Section~\ref{sec:orbiterrors}.} \label{tab:params}
\tablenotetext{a}{The preferred value for eccentricity reported in \citet{Agol2010} is $e \cos \omega = 0.000050$ (the error listed in this table is formally an error on $e \cos \omega$) but we choose $e=0$ for our default analysis. In Section~\ref{sec:orbiterrors} we show that this does not influence our results.}
\end{deluxetable}

We can measure spatial information on a planet when the light curve shows a deviation from what would be expected for an uniform flux pattern on the planet disk. We want to calculate eigencurves for these deviations, but in order to assure that a sum of our set of spherical harmonics could still create a completely uniform map, our full set includes both a positive and negative version for each spherical component (e.g. both $Y_1^1$ and $-Y_1^1$). We exclude the $Y_0^0$ (uniform) map from the PCA eignecurve analysis, since it will be part of the eventual fit to data. We use all spherical harmonics up to $l_{\rm max}=2$, resulting in a total of 16 spherical harmonic components and 16 corresponding light curves. We did also perform an analysis using $l_{\rm max}=3$ and found that the results agreed with the $l_{\rm max}=2$ analysis. For reference, \citet{Majeau2012} used $l_{\rm max}=1$ for their analysis of the HD~189733b data, while \citet{deWit2012} tested solutions up to $l_{\rm max}=3$.

We then run the set of harmonic light curves through a principal component analysis, which calculates the eigenvalues and eigenvectors of the covariance matrix of a mean-subtracted dataset. Or, to put it more plainly, PCA uses matrix math to determine new, orthogonal coordinates axes in multi-dimensional data, sorting the axes by how much variance in the data is expressed along each one. Our input dataset is the $[N_{times},N_{curves}]$ matrix of all harmonic lightcurves, while the output is a variance-ranked set of $N_{curves}$ orthogonal light curves (i.e., the eigenvectors), each of length $N_{times}$ and with an associated eigenvalue that quantifies how much of the relative variation in the input dataset can be represented by that eigenvector.\footnote{We tested both the standard covariance and singular value decomposition methods and found that they achieved identical results.} These output eigencurves, $E_n(t)$, are linear combinations of the spherical-harmonic-based light curves, $F_l^m(t)$:
\begin{equation}
E_n(t)=\sum_{l=1}^{l_{max}} \sum_{m=0}^{\pm l} \lambda_{n,l,m} F_l^m(t),
\label{eqn:eigencurve}
\end{equation} 
where $\lambda_{n,l,m}$ are the coefficients for the $n$-th eigencurve, determined from PCA. Since we input a set of 16 light curves into the PCA, we end up with a set of 16 ``eigencurves" ($n=1$ to 16). We also test three choices for $N_{times}$, for different time coverage of the planet's orbit (as described below), meaning that we use PCA to determine three different sets of eigencurves.

Each eigencurve, by construction, represents a deviation from the light curve for a uniformly bright planet, $F_0^0$, which must be included in any fit of the eigencurves to actual data:
\begin{equation}
F(t)= C_0 F_0^0(t) + \sum_{n=1}^{n_{\mathrm{max}}} C_n E_n(t) + F_{*,corr} \label{eqn:lightcurve}
\end{equation} 
where $\{C_0, \ldots ,C_{n_{\rm max}} \}$ are the fitted coefficients for each component and the value of $n_{\mathrm{max}}$ is set by the precision of the measurements. One limitation of the PCA approach being purely mathematical is that it is unaware of physical limitations. In particular, although the shape of the eigencurves are always flat during the time of secondary eclipse, they are not always zero. This motivates our inclusion of the constant term $F_{*,corr}$ in Equation~\ref{eqn:lightcurve}, to correct for non-zero eigencurve fluxes during secondary eclipse and effectively re-normalize the planet-to-star flux ratio (since a constant flux value can only be attributed to flux from the star). Lastly, we note that while the $E_n$ are orthogonal to each other by construction, they are not in general orthogonal to $F_0^0$ or $F_{*,corr}$.

\subsection{Information Content of Lightcurves}

\begin{figure}[htb]
\includegraphics[width=0.5\textwidth]{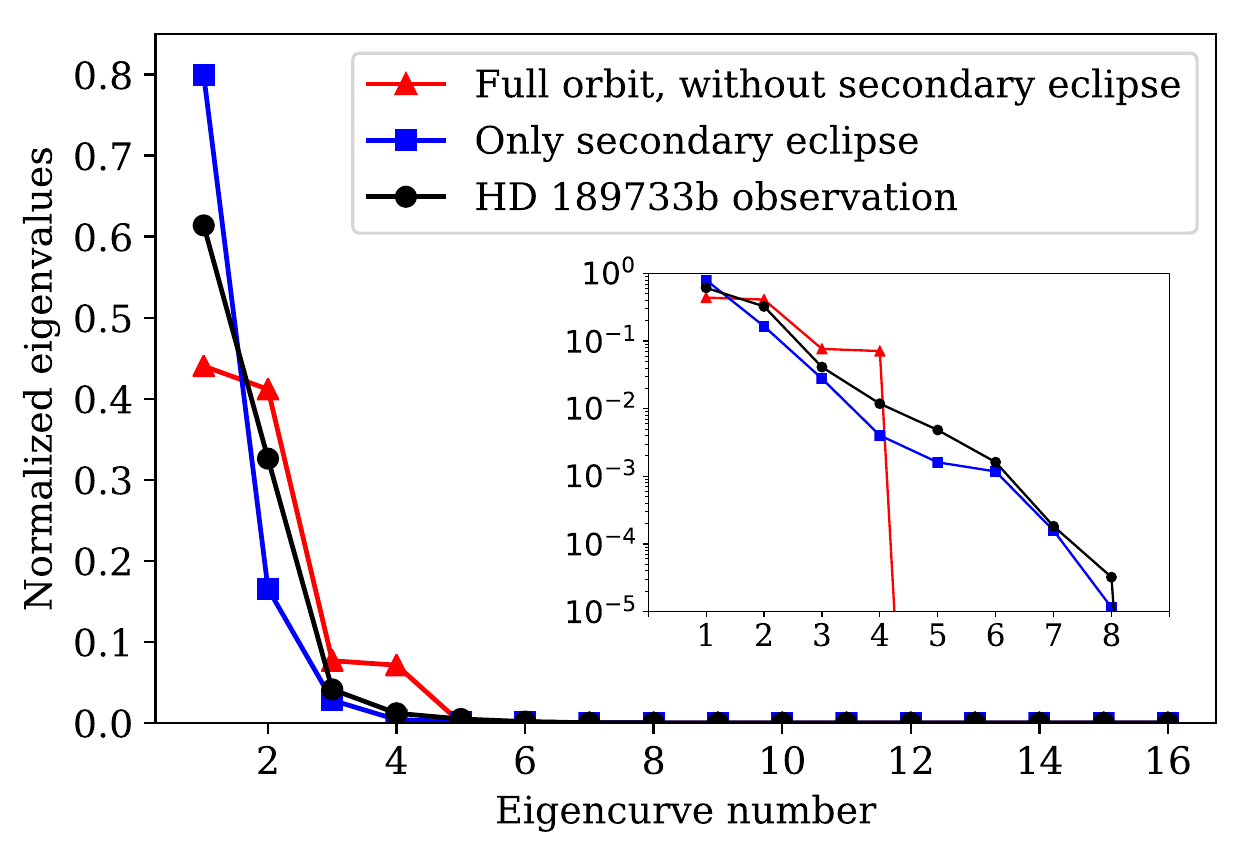}
\caption{The eigencurve power spectra for the hot Jupiter HD~189733b: the full orbit with the secondary eclipse excluded (red triangles), only the time immediately around secondary eclipse (blue squares), and the same times as the published 8 micron data for this planet (black circles). These three cases correspond to the top, middle, and bottom panel of Figure \ref{fig:eigencurves} and the top, middle, and bottom sets of maps in Figure \ref{fig:eigenmaps}. For some measurement precision, this plot informs how many pieces of spatial information can be measured, as well as how much each component could contribute to the observation. The inset shows the same values on a logarithmic scale, with values off the bottom of the plot computationally equal to zero. There are only four non-zero components for the full orbit case because we only include spherical harmonics up to $l=2$ in our analysis.} \label{fig:default_latent}
\end{figure}

Figure~\ref{fig:default_latent} shows the normalized eigenvalues of the covariance matrix, sorted in order of which components produce the largest variance, for our three different $N_{times}$ PCA calculations. In other words, this plot shows how strongly each eigencurve can contribute to a total light curve signal. The eigencurves corresponding to those largest eigenvalues are then the mathematically ideal curves to use in fitting the data. We compare three different cases: 1) an analysis in which we include light curves for one full orbit of the planet but exclude the secondary eclipse, 2) an analysis using only a fraction of the planet's orbit, covering $\sim$0.05 of an orbital period before and after secondary eclipse, and 3) an analysis for the same time coverage as our data for HD 189733b, roughly a quarter of an orbit. In the first two cases the time sampling is identical (500 points in time for one orbit), while the time sampling for the third is set to match that of the actual observations. The data are described below (Section~\ref{sec:data} and Figure~\ref{fig:HD189_lc}); they are an uneven combination of separate observations, with a total of $N_{times}=880$, but the time sampling is about 7 times higher for the period around secondary eclipse. 

\begin{figure*}[htb]
\begin{center}
\includegraphics[width=0.65\textwidth]{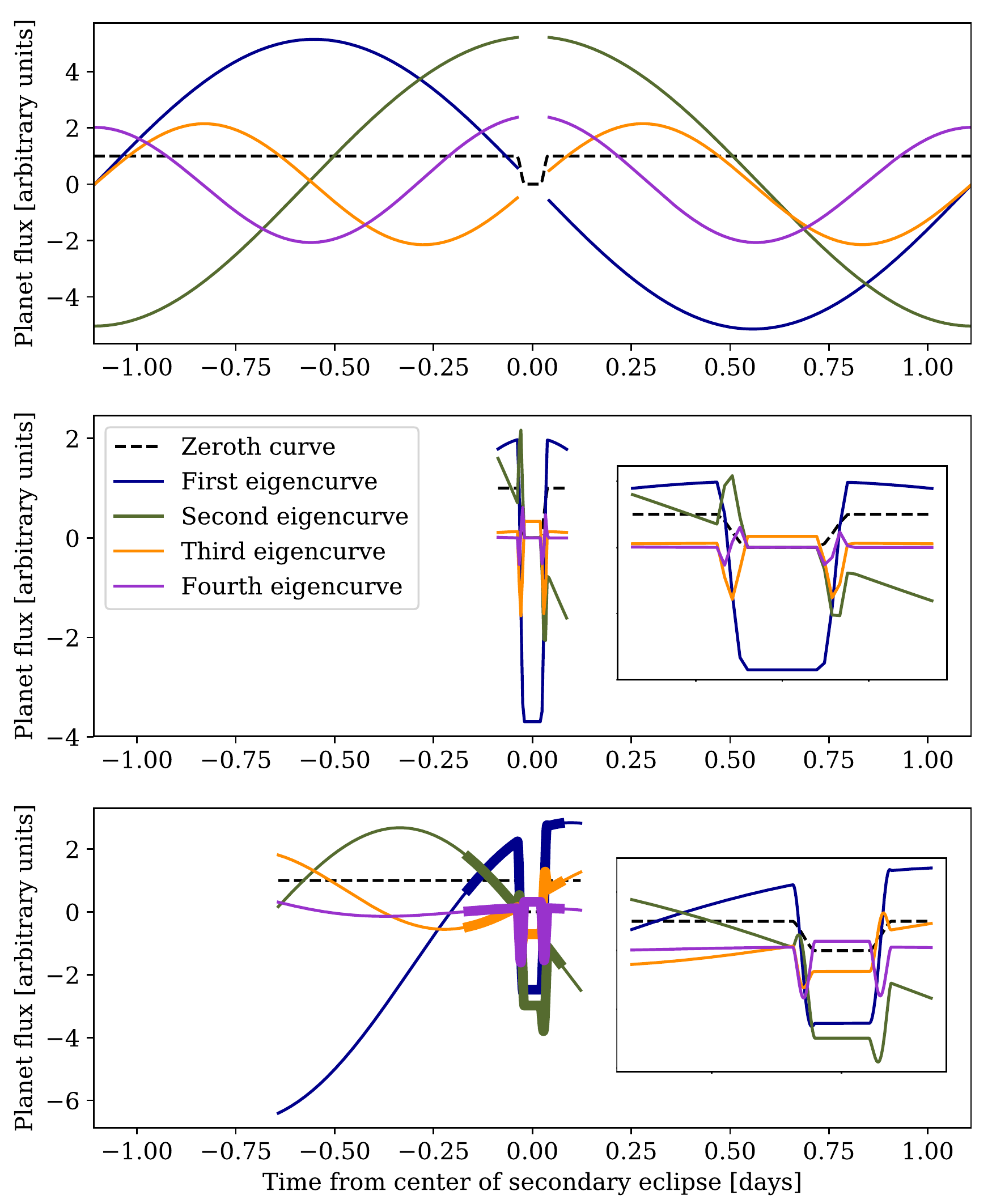}
\caption{The uniform planet lightcurve, $F_0^0(t)$, and first four eigencurves, $E_{n=1,2,3,4}(t)$, for three different observational scenarios of the HD~189733b hot Jupiter system. \textit{Top:} a full-orbit phase curve without secondary eclipse, \textit{middle:} an observation of just the time around secondary eclipse, and \textit{bottom:} the Spitzer 8~micron observations of the hot Jupiter HD~189733b. Note that for the observation shown in the bottom panel the data are unevenly sampled in time: the times shown as a thicker line (also the range plotted in the inset) correspond to the seven eclipse measurements. 
We also note that the time sampling we have chosen means there are only three points each during ingress and egress in the middle plot; however this is sufficient to accurately retrieve the low order eigencurves, as proven in comparison to tests with higher time resolution. The orbital phases at which we observe a planet determine which lightcurve components---and therefore spatial information---we are sensitive to.} \label{fig:eigencurves}
\end{center}
\end{figure*}

While Figure~\ref{fig:default_latent} informs how many pieces of spatial information can be retrieved for some measurement precision, we can also investigate the information-sorted eigencurves in order to determine what those pieces of information are. In Figure~\ref{fig:eigencurves} we show the first four most-informative eigencurves ($n=1,2,3,4$) for each of the three cases. We see in Figure \ref{fig:default_latent} that a full-orbit observation retrieves paired pieces of information and Figure~\ref{fig:eigencurves} shows that these correspond to eigencurves that have a form like paired sine and cosine functions, followed by paired curves like $\sin 2\phi$ and $\cos 2\phi$ functions. In contrast, mapping using just the time around secondary eclipse only has one main piece of information primarily available, with the number of additional components strongly dependent on the level of noise in the data. 

The presence of only four non-zero eigenvalues for the full phase curve case in Figure~\ref{fig:default_latent} is the result of our choice to only include spherical harmonics up to $l_{max}=2$ in our analysis. If we had exactly equatorial viewing geometry for this planet, odd modes (above $m=1$) would be completely invisible in thermal orbital phase curves \citep{Cowan2008}. Since HD~189733b is not in an exactly edge-on orbit, odd modes with north-south asymmetry could be observable in thermal phase curves \citep[starting with $Y_4^3$;][]{Cowan2013a}, but we have not included such high-order spherical harmonics in our analysis. A similar argument explains why there are 8 non-zero eigenvalues for the cases including the eclipse: there are 9 spherical harmonics up to $l_{max}=2$, but we don't use the zeroth lightcurve in our PCA.  We use 16 lightcurves, but half of them are negative versions of, and hence proportional to, the 8 usual harmonic lightcurves.  Indeed, repeating the analysis with $l_{max}=3$ produced 15 non-zero eigenvalues, as expected.

\section{Eigenmaps} \label{sec:eigenmaps}

\begin{figure*}[htb]
\begin{center}
\includegraphics[width=\textwidth]{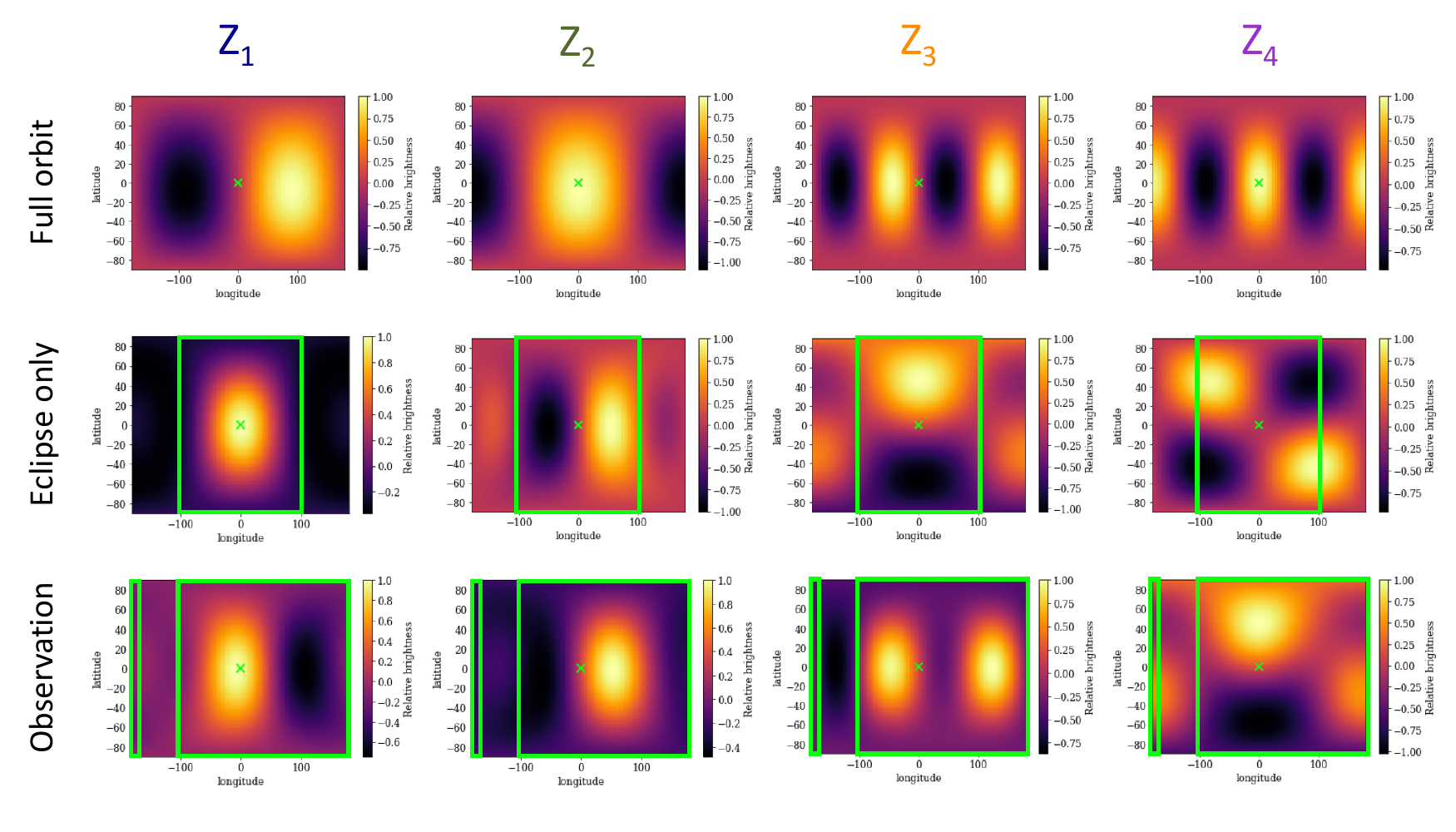}
\end{center}
\caption{The eigenmaps corresponding to the eigencurves presented in Figure \ref{fig:eigencurves}, with the point facing the star marked as a green ``X" and arbitrary flux normalization. From left to right are the first through fourth eigenmaps, $Z_{n=1,2,3,4}(\theta,\phi)$. When there is only phase curve information available (top row), we are sensitive to large-scale longitudinal patterns that are close to $m=\pm l$ spherical harmonics. For secondary eclipse mapping (middle row), we retrieve large-scale dayside flux gradients but are relatively insensitive to flux patterns on the nightside. The 8~micron \emph{Spitzer} observations of HD~189733b (bottom row) covers about a quarter of an orbit and is more highly sampled near secondary eclipse, producing eigenmaps that are a hybrid of the first two rows. The green boxes show the regions of the planet that are visible to the observer at some point in the observations; for the full-orbit scenario, the entire planet can be mapped, while in the two bottom rows, some regions never face the observer. }
\label{fig:eigenmaps}
\end{figure*}

Since each eigencurve is composed of a linear combination of light curves calculated from spherical harmonic maps, the same coefficients that make up the eigencurve can be used to calculate the corresponding eigenmap, $Z_n(\theta, \phi)$:
\begin{equation}
Z_n(\theta, \phi) = \sum_{l=1}^{l_{max}} \sum_{m=0}^{\pm l} \lambda_{n,l,m} Y^m_l(\theta,\phi)
\label{eqn:eigenmap}
\end{equation}
where $\theta$ is co-latitude and $\phi$ is longitude. In Figure~\ref{fig:eigenmaps} we show the first four eigenmaps ($n=1,2,3,4$) for each of our three observational cases (full-orbit, eclipse-only, and actual measurement), corresponding to the sets of eigencurves in Figure~\ref{fig:eigencurves}. The retrieved map for the planet will be given by:
\begin{equation}
Z_p(\theta,\phi)= C_0 Y_0^0(\theta,\phi) + \sum_{n=1}^{n_{\mathrm{max}}} C_n Z_n(\theta,\phi), 
\label{eqn:finalmap}
\end{equation}
where $C_n$ are the same coefficients from Equation~\ref{eqn:lightcurve}, determined by a fit to light curve data. (We discuss how the $F_{*,corr}$ term in Equation \ref{eqn:lightcurve} is explicitly treated when calculating units for planetary flux in Section \ref{sec:data}.)

Eigenmaps, much like spherical harmonics, are global basis maps \citep[as opposed to pixel basis maps, which are local;][]{Cowan_Fujii2017}. As such, fitting data with eigencurves will necessarily produce a map with some ``information'' about regions of the planet that were never visible. Since local basis maps in practice require smoothing \citep[e.g.,][]{Knutson2007b,Majeau2012}, they also provide a glimpse of unseen regions; this seeming paradox is due to the assumption of smoothness in the brightness map of exoplanets.  That said, another advantage of eigenmaps over spherical harmonics is that the former are relatively featureless in the regions that are least probed by the data, e.g., the first eigenmap in the eclipse-only scenario has a nearly uniform nightside.

Note that these are the orthogonal maps we are most sensitive to, given certain observations. Whether a given planet actually exhibits these modes in its brightness map is up to Nature. For example, as we discuss below, the eclipse observations of HD~189733b are most sensitive to the central concentration of flux on the planet's dayside, but the data for this planet favor a relatively uniform dayside brightness. Nonetheless, the eigenmaps can best constrain the flux pattern on the planet (a lack of power in a mode is still informative) and if the planet exhibits structure too detailed to retrieve with low order eigenmaps, our analysis shows that we either need better photometric precision, or those details may just be inaccessible to mapping. 

The predictions of specific spatial patterns from atmospheric models can be tested in order to determine what information can be retrieved in a map. This would be done by taking an inner product of the flux predictions with the eigenmaps, to determine how much power is in each component. This should be the same, but mathematically faster, than calculating predicted flux curves and running a retrieval on simulated data. (Although the later approach has the benefit of also testing measurement precision.) Since the eigenmaps for phase curve versus eclipse-only observations are different, as we are about to see, this also would help to determine which type of measurement can best retrieve the predicted spatial information.

\subsection{Full-Orbit}
Focusing on the full-orbit case first, we can see that the sine-like and cosine-like eigencurves we recognized in Figure~\ref{sec:eigencurves} intuitively correspond to their eigenmaps. The first pieces of information we can learn about the planet are the largest, hemispheric differences and the first two components (of almost equal information content, see Figure~\ref{fig:default_latent}) are just the 90\degr~phase-shifted versions of this pattern. Similarly, the next two eigenmaps provide information about the next higher order spatial variation in longitude and are also of about equal information content. In all cases the slightly non-equatorial viewing geometry of the planet induces a very small break in the north-south symmetry of the eigenmaps, as one hemisphere is slightly more directly viewed.

\subsection{Eclipse-Only}
Our results for the eclipse-only case paint a different picture, but one that is also intuitive (and in all cases show the same slight break from north-south symmetry seen in the full lightcurve results). The first eigenmap, which contains almost all of the available information, characterizes the largest-scale flux gradient present on the planet's day side (the region resolved by secondary eclipse). Since the eigencurve and eigenmap can just as easily be multiplied by a positive or negative coefficent, this feature could be a flux increase or decrease near the substellar point. The next two components provide information on any east-west or north-south shift of the flux pattern (and again, the sign of the coefficient that would multiply this map determines which direction the shift would be), and the fourth component describes a rotation of the flux pattern. Note that all eigenmaps have less variation in structure on the night side of the planet, since this region is not observed in an eclipse-only measurement.

As shown in Figure~\ref{fig:default_latent}, we find that the $n=2$ component in Figure~\ref{fig:eigenmaps} contributes significantly more to the information content of the lightcurve than the third component, even though these correspond to geometrically similar flux shifts in orthogonal directions.  This is due to the out-of-eclipse baseline that we have included as extra information before and after eclipse, to be consistent with the way that actual secondary eclipse measurements are performed. Although this is only a small fraction of the full orbital phase curve, it is enough for the data to preferentially inform the longitudinal pattern of the planet, instead of its latitudinal structure.\footnote{The ability to use the time just before and after secondary eclipse to measure a small amount of longitudinal information was demonstrated by \citet{Wong2014}, who found a 4-$\sigma$ difference from a flat phase curve, for their combined 12 secondary eclipse measurements of the eccentric hot Jupiter XO-3b.}

We did calculate eclipse-only eigencurves with the inclination of the system set to 90 degrees (i.e., for an exactly edge-on system). In this special case, where the eclipse impact parameter is zero and the motion of the stellar limb across the planet disk is entirely longitudinal, the information about the North--South asymmetry of the planet is no longer accessible. However, we find that we can measure latitudinal information for systems that are even just a little bit away from edge-on \citep[cf.\ Figure~7 of][]{Cowan2013a}; non-zero eigenmaps with north-south asymmetry exist for a test case with $i=89.9$ degrees, albeit appearing later in the eignevalue-sorted series than for HD~189733b's actual inclination. (Maps similar to the $Z_3$ and $Z_4$ ones in Figure \ref{fig:eigenmaps} respectively become the 7th and 6th eigenmaps for the $i=89.9$ degrees case.) At the level of precision with which we know the impact parameter of HD~189733b, the latitudinal information is not affected (Section \ref{sec:orbiterrors}).

\subsection{Partial Phase Curve}
Finally, we also present in Figure~\ref{fig:eigenmaps} the first four eigenmaps for the time sampling of the mapping-quality 8~micron data for the planet HD 189733b. Unsurprisingly, this scenario is an intermediate cross of the full-orbit and eclipse-only cases. The first two maps in particular contain much more information on the dayside and eastern parts of the planet than the other regions, with these being the longitudes primarily viewed for an observation that starts after transit and runs to a little after secondary eclipse. The first three maps together are similar to the first three maps for the full-orbit case, since they primarily provide information on the longitude brightness structure. The relative contribution of the secondary eclipse to the full data set means that latitudinal information becomes available in the $n=4$ eigenmap, which is later than for the eclipse-only case ($n=3$), but this information is completely inaccessible in the full-orbit case because it lacks the information from secondary eclipse.

\section{Orbital uncertainties and their impact on retrieved planet maps} \label{sec:orbiterrors}

Each planetary system has unique orbital parameters and hence unique harmonic lightcurves, eigencurves, and eigenmaps. However, uncertainties in those orbital parameters are directly degenerate with uncertainty in the planet's brightness distribution. This was recognized as early as \citet{Williams2006}, who noted the link between a planet's eccentricity and east-west shifts in its brightness pattern. More recently, this problem was thoroughly explored by \citet{deWit2012}, who identified a multi-parameter degeneracy between eccentricity, impact parameter, stellar density\footnote{The star's mass controls the planet's orbit and, together with the star's size, this controls how long the planet spends in transit or eclipse.}, and the planet's brightness distribution. Including orbital information from radial velocity measurements, and not just relying on the phase curve alone, can significantly reduce uncertainties in the planet map \citep{deWit2012}.

Our method allows us to directly test how much, if at all, uncertainties in a planet's orbital parameters may limit our ability to retrieve a robust map. We do this by creating a larger set of harmonic curves to feed into the PCA; instead of just a set of lightcurves corresponding to spherical harmonics at the preferred orbital values, we include two extra sets of lightcurves, which are realizations of the spherical harmonic maps with an orbital parameter set to $\pm1\sigma$ of its preferred value. The PCA then calculates eigencurves that are linear combinations of both the spherical harmonics and the orbital parameters. If an orbital parameter has little influence on an eigencurve, then there should be little difference between the coefficients assigned to the different orbital realizations of any given spherical harmonic ($\lambda_{n,l,m}$ in Equations \ref{eqn:eigencurve} and \ref{eqn:eigenmap}, but now with an extra index for the orbital realization).

We apply this test to the HD~189733 system, in order to evaluate our orbital-mapping uncertainties before analyzing the data. Using the uncertainties from Table~\ref{tab:params}, we check the impact of the $1\sigma$ errors on orbital inclination, eccentricity\footnote{Formally, the error on eccentricity is on $e \cos \omega$, but SPIDERMAN treats $e$ and $\omega$ as separate parameters. We choose to set $\omega=0$ and vary $e$, but tests of $\omega$ produce much the same result.}, and scaled semi-major axis, which are analogous to the degenerate parameters identified by \citet{deWit2012}. We ran this sensitivity analysis for two time ranges: one that mimicked all possible information (a fully sampled orbit, including secondary eclipse, with 500 points evenly spread in time) and one for the actual time sampling of our data.

We find that in both the full orbit and partial orbit scenarios, and for all three tested orbital parameters, there is a $\lesssim$1\% difference between the coefficients assigned to the different orbital realizations for the first five eigencurves. In many instances the differences are substantially smaller; for the case of our actual data, the eccentricity and scaled semi-major axis each influence the first two eigencurves by less than 0.1\%. In fact, up to the maximum retrievable pieces of information ($n_{\mathrm{max}}=8$, set by the last eigencurve to have a non-zero eigenvalue, see Figure~\ref{fig:default_latent}), the differences between orbital realizations never gets above $\sim10$\%.

The orbital uncertainties for HD~189733b are sufficiently small that they do not impact the low-order eigencurves that can be constrained by the extant photometry or, in other words, we find that orbital uncertainties do not limit our ability to map this planet. It should be noted that HD~189733b is among the best characterized exoplanets and is also the only exoplanet for which phase+eclipse mapping has been performed.  Indeed, we generally expect that the planets benefiting from the most precise photometry will also have the best orbital constraints, for the simple reason that the host star is probably bright. 
If researchers want to map a planet with poorly constrained orbital parameters, our eigencurve method can be used to determine whether those uncertainties will disallow the retrieval of a robust map. If the orbital errors are too large, this same technique can be used to estimate how much better constrained they would need to be in order to successfully map the planet, perhaps motivating additional radial velocity observations.

\section{Application: a map of HD 189733\lowercase{b}} \label{sec:data}

\begin{figure*}[htb]
\begin{center}
\includegraphics[width=\linewidth]{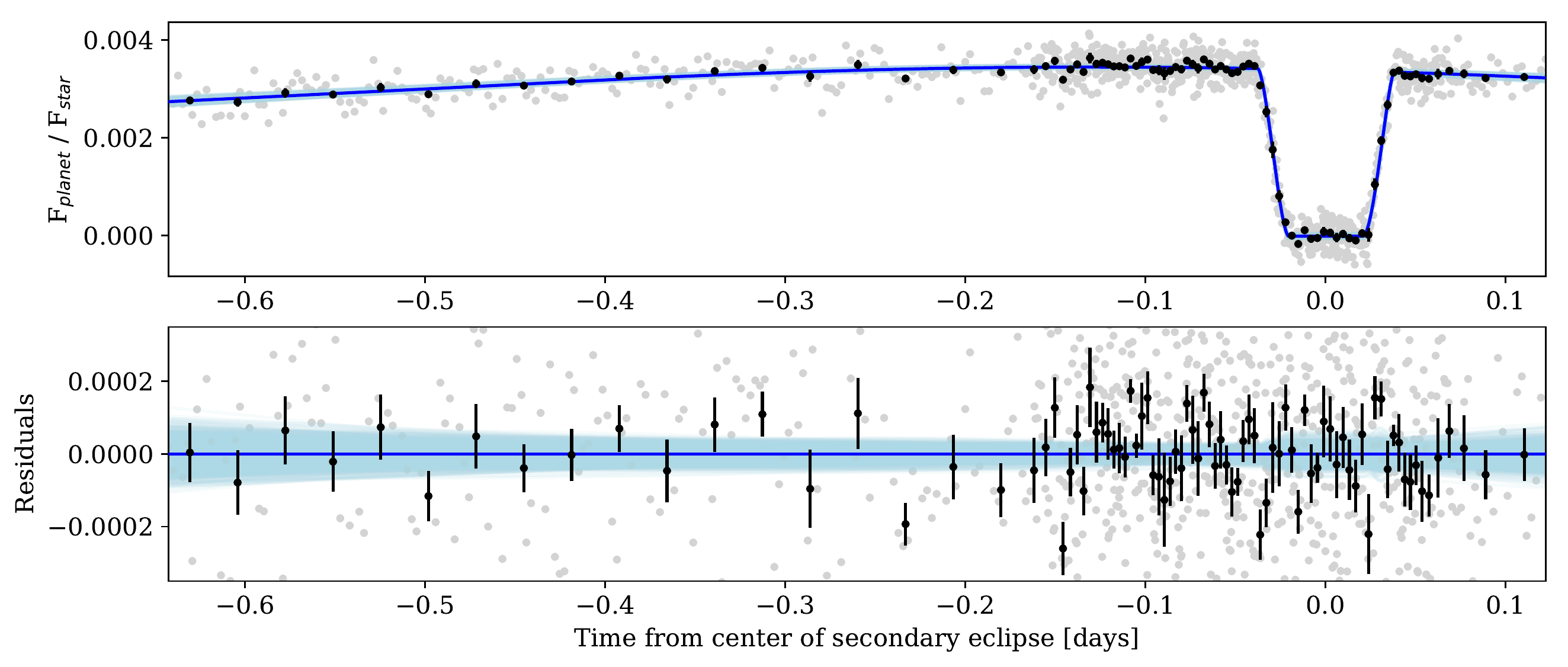}
\end{center}
\caption{Our preferred fit ($n_{max}=2$) to the 8~$\mu$m photometry of the hot Jupiter HD~189733b (top) and residuals (bottom). The black points are the data (gray points) binned in sets of 10, with the error bar giving the standard deviation of those points divided by $\sqrt[]{10}$. The dark blue line shows the preferred fit, while the light blue lines are 1000 draws from the MCMC chain, a visual demonstration of the uncertainty in the fit. The scale chosen for the residuals plot leaves some of the gray data points off of the range shown, but lets the reader more carefully visualize the binned data compared to the model fits. \label{fig:HD189_lc}}
\end{figure*}

In order to demonstrate our eigencurve/eigenmap method in practice, we apply it to the one exoplanet for which phase and eclipse mapping has been performed: the hot Jupiter HD~189733b \citep{deWit2012,Majeau2012}. We use the same reduced and detector-corrected Spitzer Space Telescope observations of this planet as in \citet{Majeau2012}, which is a combination of eclipses from \citet{Agol2010} and phase curve data originally from \citet{Knutson2007b}, as re-reduced and corrected in \citet{Agol2010}. The original phase measurements spanned roughly half an orbit, from transit to eclipse.  But due to problems correcting for detector systematics at the start of the observations, the trustworthy phase observations only span about a quarter of an orbit---these are the observations used in \citet{Agol2010}, \citet{Majeau2012}, and in the current study (as shown in Figure~\ref{fig:HD189_lc}).

Using the eigencurves calculated above (Figure~\ref{fig:eigencurves}), we experiment with using varying numbers of components in fits of Equation \ref{eqn:lightcurve} to the data. For the $F_0^0(t)$ function in Equation~\ref{eqn:lightcurve} (for $Y_0^0$, the uniform disk), we use the light curve calculated by SPIDERMAN with the HD~189733b system geometry, using fixed values equal to those in Table \ref{tab:params}. Our fit to the data involves finding preferred values for $F_{*,corr}$, $C_0$, and the $C_n$ coefficients corresponding to whichever $E_n$ eigencurves we choose to include. We use a Markov chain Monte Carlo \citep[MCMC, namely emcee;][]{ForemanMackey2013}, with uniform priors and initialized by optimizing the likelihood function. For each fit we used 300 walkers and 500 steps. By examining the time series of walkers for each parameter in the sampler chain we estimate that the ``burn-in'' period for our fits is less than 100 steps (after which there were no significant changes in the walkers' exploration of each parameter) and so for each fit we discarded the first 100 steps. For each MCMC fit we identify the parameter set in the sampler chain with the maximum likelihood and use those coefficients as our best fit values. From the distributions in each sampler chain we calculate the differences between the 84th and 50th percentiles and between the 50th and 16th percentiles as the upper and lower uncertainties on the coefficients. All of these values are reported in Table~\ref{tab:fits}, for a subset of the fits we tested.

\begin{deluxetable*}{l|cc|ccccccc}[hbt]
\tablecaption{Best fit coefficients and uncertainties (all values divided by $1\times 10^{-4}$) and goodness-of-fit parameters.}
\tablehead{\colhead{Model fit} & \colhead{$\chi^2$} & \colhead{BIC} & \colhead{$F_{*,corr}$} & \colhead{$C_0$} & \colhead{$C_1$} & \colhead{$C_2$} & \colhead{$C_3$} & \colhead{$C_4$} & \colhead{$C_5$}} 
\startdata
$n_{max}=1$ & 1054 & 1074 & 1.04$^{+0.23}_{-0.22}$ & 10.205$^{+0.082}_{-0.083}$ & 0.522$^{+0.036}_{-0.036}$ & -- & -- & -- & -- \\
$\mathbf{n_{max}=2}$ & \textbf{942.4} & \textbf{969.6} & \textbf{4.52}$^{+0.39}_{-0.40}$ & \textbf{8.84}$^{+0.16}_{-0.15}$ & \textbf{0.790}$^{+0.045}_{-0.045}$ & \textbf{0.906}$^{+0.085}_{-0.086}$ & -- & -- & -- \\
$n_{max}=3$ & 938.8 & 972.7 & 9.9$^{+3.0}_{-2.9}$ & 6.7$^{+1.1}_{-1.2}$ & 1.21$^{+0.23}_{-0.23}$ & 1.93$^{+0.57}_{-0.55}$ & 1.79$^{+0.98}_{-0.95}$ & -- & -- \\
$n_{max}=4$ & 936.4 & 977.1 & 10.7$^{+3.0}_{-3.0}$ & 6.4$^{+1.2}_{-1.2}$ & 1.27$^{+0.23}_{-0.24}$ & 2.07$^{+0.56}_{-0.58}$ & 2.03$^{+0.98}_{-1.00}$ & -0.34$^{+0.24}_{-0.24}$ & -- \\
$n_{max}=5$ & 936.4 & 984.0 & 10.3$^{+3.1}_{-3.1}$ & 6.6$^{+1.2}_{-1.2}$ & 1.23$^{+0.24}_{-0.23}$ & 1.99$^{+0.58}_{-0.58}$ & 1.9$^{+1.0}_{-1.0}$ & -0.33$^{+0.24}_{-0.23}$ & -0.14$^{+0.38}_{-0.39}$ \\ \hline
non-seq 2a & 951.0 & 978.1 & -0.23$^{+0.26}_{-0.25}$ & 10.705$^{+0.094}_{-0.096}$ & 0.422$^{+0.038}_{-0.038}$ & -- & -1.52$^{+0.14}_{-0.15}$ & -- & -- \\
non-seq 2b & 966.6 & 993.7 & -5.44$^{+0.43}_{-0.43}$ & 12.75$^{+0.17}_{-0.16}$ & -- & -0.966$^{+0.091}_{-0.094}$ & -3.22$^{+0.18}_{-0.19}$ & -- & -- \\
non-seq 3  & 941.0 & 974.9 & 4.58$^{+0.40}_{-0.39}$ & 8.82$^{+0.15}_{-0.15}$ & 0.792$^{+0.045}_{-0.044}$ & 0.919$^{+0.086}_{-0.086}$ & -- & -0.29$^{+0.23}_{-0.24}$ & -- \\
\enddata
\tablecomments{Each row in this table gives information for a different model we tried fitting to the data, with the fit's $\chi^2$, Bayesian Information Criterion (BIC), and values for each coefficient included in the fit. We report the parameter set in each MCMC chain with the maximum likelihood and uncertainties that come from the $\pm34$ percentile distributions of each parameter. For reference, there are 880 data points and anywhere from 3 to 7 free parameters, depending on the model fit. Our preferred fit is the $n_{max}=2$ model.}
\label{tab:fits}
\end{deluxetable*}

We tested fits using models with an increasing number of eigencurves, from those most to least easily detected by the observation, as quantified by their eigenvalues (Figure \ref{fig:default_latent}). These sequential models are listed in Table \ref{tab:fits} as $n_{max}=1$ through 5. However, the observability of any given eigenmap is purely mathematical and may or may not correspond to a realized mode in the actual physical structure of the planet. As such, we also tested fits using models with non-sequential combinations of eigencurves; the three best-fitting of those tests are also shown in Table \ref{tab:fits}, as ``non-seq'' with a number indicating how many eigencurves were included (in addition to $C_0$ and $F_{*,corr}$, both of which must be used in all fits).

Table \ref{tab:fits} reports the $\chi^2$ values for each model fit and, as should be the case for converged fits, we find that every time we add a new term to the $n_{max}$ fits the $\chi^2$ decreases, as the model is better able to fit the data. This is also true in comparing the $n_{max}=2$ model to the ``non-seq 3" model, which has one additional component. Interestingly, the non-sequential model fits always have worse $\chi^2$ values than $n_{max}$ models with an equal number of parameters. This is maybe not surprising since none of the first four eigencurves exhibit zero power (it is not until $C_5$ that a coefficient value is consistent with zero), meaning that the physical pattern on the planet is well represented by the maximally informative eigenmaps.

Even though we expect $\chi^2$ values to decrease as more parameters are included in a fit, the criterion for a statistically significant improvement in the fit is that the reduction in $\chi^2$ is greater than $\sim \sqrt[]{\chi^2}$. As such, the $n_{max}=3$ fit is not significantly improved over the $n_{max}=2$ fit. This statistical statement is further supported by comparing the Bayesian Information Criterion (BIC) values for these fits, which show that the improved agreement between the data and the model is not enough to warrant adding another eigencurve in the fit. The difference between the BIC values for the $n_{max}=2$ and $n_{max}=3$ fits shows positive (albeit not strong) evidence for the $n_{max}=2$ fit.

The reader may notice that the $n_{max} \geq 3$ solutions seem to have converged on a set of coefficients significantly different from, and with larger associated errors than, the $n_{max}=2$ solution. Upon inspection, we discovered that those MCMC fits all show strong degeneracies between the fit coefficients, resulting in the parameter space being sampled much more broadly, as the significance of the particular value of any coefficient is lost in comparison to its combined value with other coefficients. This means that those fits have not converged on a more correct solution, but rather exhibit a highly degenerate model set-up. This is another piece of evidence in favor of the $n_{max}=2$ solution.

\begin{figure}[htb]
\includegraphics[width=\linewidth]{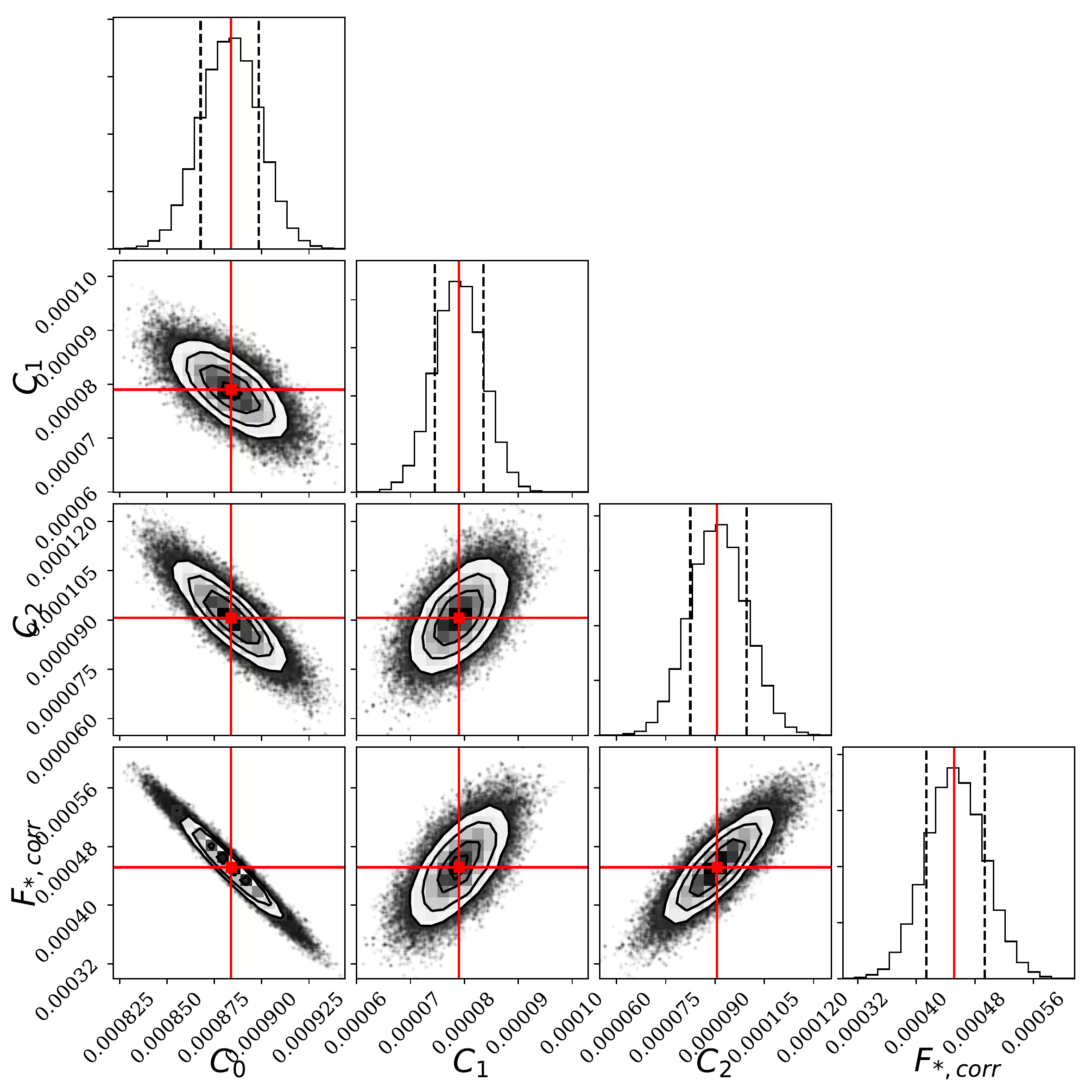}
\caption{The corner plot from the fit using $n_{max}=2$. The red squares and lines mark the values of the coefficients in the maximum likelihood solution. The black dashed lines mark the 16th and 84th percentiles in each parameter distribution. The components fit by the $C_1$ and $C_2$ coefficients are mathematically orthogonal to each other (by design), but this is not the case between other combinations of components. We see some correlation between the $C_0$ and $F_{*,corr}$ coefficients, and weaker correlation between other pairs of coefficients.} \label{fig:corner}
\end{figure}

We present our preferred, $n_{\mathrm{max}}=2$, fit to the data in Figure~\ref{fig:HD189_lc}, as well as the residuals. The dark blue line is the curve with the maximum likelihood from the MCMC fit. We also plot light blue curves corresponding to 1000 random samples of sets of $[C_0,C_1,C_2,F_{*,corr}]$ from the MCMC results, to visualize the uncertainty. Figure \ref{fig:corner} shows the projections of the MCMC sampler chain onto the planes of each combination of fit coefficients, with the maximum likelihood coefficient values sitting nicely in the center of each distribution. We see some correlation between the $F_{*,corr}$ and $C_0$ coefficients, and to a lesser extent between those coefficients and $C_1$ and $C_2$, which then induces the slight correlation seen between the mathematically orthogonal $C_1$ and $C_2$ coefficients.

\begin{figure}[htb]
\includegraphics[width=\linewidth]{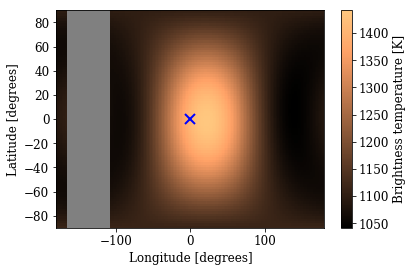}
\caption{The 8~$\mu$m brightness temperature map of HD~189733b, as calculated from the preferred fit of our eigencurve mapping technique. The point that permanently faces the star is marked with a blue ``X". The region of the planet that faces away from the telescope during the observation is obscured by a solid gray box. This fit only includes eigenmaps up to $n_{\mathrm{max}}=2$ (the two leftmost panels in the bottom row of Figure \ref{fig:eigenmaps}), in addition to a uniform brightness component, meaning that the longitude of the brightest region is well constrained but there is no information about any north-south asymmetry available.} \label{fig:HD189_map}
\end{figure}

We use the maximum likelihood set of coefficients in Equations~\ref{eqn:eigenmap} and \ref{eqn:finalmap} to calculate a corresponding preferred fit planet flux map. We convert this to brightness temperature (described in Appendix~\ref{sec:tbright}) and apply our stellar flux correction factor (Appendix~\ref{sec:fstar}), arriving at the preferred planet brightness temperature map we show in Figure~\ref{fig:HD189_map}. A gray rectangle covers the region of the planet that was out of view for the entire combined observation.
 
\begin{figure}[htb]
\includegraphics[width=\linewidth]{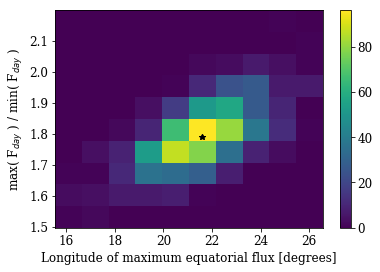}
\caption{A two-dimensional histogram of the planet's physical parameters, calculated from mapped realizations of 1000 samples drawn from our MCMC fit. The horizontal axis gives the longitude (east of the substellar point) where the planet is brightest, while the vertical axis is the maximum dayside flux divided by the minimum dayside flux. The black star (with infinite space-time curvature) marks the values from the maximum likelihood solution. In addition to producing a preferred retrieved map (Figure \ref{fig:HD189_map}), we can directly calculate uncertainties on map properties from the uncertainties in fitted coefficients.} \label{fig:observables}
\end{figure}

We translate uncertainties in the light curve fit to uncertainties in the retrieved map, using the same linear math outlined above. This allows us to identify which regions of the planet are more or less reliably mapped. From Figure~\ref{fig:eigenmaps} we can see that the spatial patterns of the first two eigenmaps are primarily informing us of the longitude of the brightest region of the atmosphere, through the signs and relative amplitudes of $C_1$ and $C_2$. These components also influence the flux gradients across the planet. In order to characterize how well we have measured these physical properties, we draw 1000 samples from our MCMC fit, calculate the longitude of maximum flux (along the equator), as well as the maximum dayside flux relative to the minimum dayside flux.\footnote{We calculate this only for the dayside as the observations cannot map the entire nightside.}
We show the distribution of these parameters in Figure~\ref{fig:observables}. From the means and standard deviations, we calculate the equatorial longitude of maximum planet flux to be $21.6 \pm 1.6$ degrees east of the substellar point 
and the dayside flux contrast to be $1.799 \pm 0.085$.

Our result for the longitude of maximum flux is remarkably consistent with the analysis of \citet{Majeau2012}, who calculated the longitude at $21.8\pm1.5$ degrees east of the substellar point. \citet{deWit2012} were cautious not to quote a particular value for the location of the brightest region of the planet, instead emphasizing that different assumptions about the brightness distribution resulted in different answers, but found solutions with longitudes ranging from $\sim10-30$ degrees east (see their Figure 16 for solutions that incorporate radial velocity data for the orbital parameters). We can also roughly compare our retrieved flux contrasts with these previous results. Figure 16 of \citet{deWit2012} shows a maximum dayside flux that appears to be about 1.7 times the minimum flux; this agrees nicely with our result of $1.799 \pm 0.085$. Figure 4 of \citet{Majeau2012} shows a minimum global flux that is about 30\% as bright as the maximum; we find this ratio to be about 50\%, but since the night side cannot be measured as well by this partial-orbit observation, this value cannot be as definitive as the dayside flux contrast.

\begin{figure}[htb]
\includegraphics[width=\linewidth]{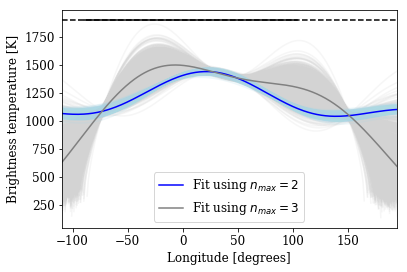}
\caption{Brightness temperature as a function of longitude along the equator, for the $n_{\mathrm{max}}=2$ and $n_{\mathrm{max}}=3$ fits, showing the maximum likelihood solutions as darker lines and 1000 draws from the MCMC samples as lighter lines. We only plot the longitudes that were measurable during the observation; the black solid line is for longitudes directly in view (or mapped during ingress+egress) and the dashed line includes those up to the planet's limb. This comparison demonstrates that the cost of adding extra information is to create very large uncertainties (including encompassing non-physical solutions).} \label{fig:lon_profiles}
\end{figure}

Finally, we also analyze the reliability of our retrieved flux map, as a function of longitude, by focusing on the equatorial profile. In Figure~\ref{fig:lon_profiles} we show two sets of profiles, for both our preferred $n_{\mathrm{max}}=2$ fit and the slightly less favored $n_{\mathrm{max}}=3$ fit. The thick dark lines are the profiles calculated from the maximum likelihood set of coefficients from the MCMC fits, while the thin lines are 1000 randomly drawn samples from each fit. This comparison between the $n_{\mathrm{max}}=2$ and $n_{\mathrm{max}}=3$ fits clearly demonstrates why the BIC preferred the former. While the extra information in the $n_{\mathrm{max}}=3$ fit allows for a more detailed flux structure, this comes at the significant cost of much larger uncertainties. Note that this model suggests a \emph{westward} hotspot offset and a large day--night contrast. In fact, many of the $n_{\mathrm{max}}=3$ realizations favor non-physical conditions on the planet's night side, with fluxes (and brightness temperatures) dropping below zero.\footnote{See Section~\ref{sec:history} and/or \citet{Keating2017} for a discussion of the problem of non-physical solutions.} At the longitudes where the $n_{\mathrm{max}}=3$ fit shows smaller uncertainty, we see that these values are in agreement with the solution already provided by the $n_{\mathrm{max}}=2$ fit, as we should expect from the orthogonality of the eigencurves. 

\section{Summary} \label{sec:summary}

In anticipation of the launch of JWST, which will have sufficient precision to enable eclipse-only and combined eclipse+phase mapping for many bright transiting planets, we have introduced a new method to calculate mathematically ideal basis functions for fitting the data. We promote the use of principal component analysis, for the specific timing of an observation, to calculate an orthogonal set of ``eigencurves", with corresponding ``eigenmaps". We use light curves from spherical harmonic maps \citep[conviently calculated by the SPIDERMAN code,][]{Louden2018} as input for our PCA, but in principle any complete set of light curves would work. Since the eigencurves and eigenmaps are the same regardless of the input lightcurves, the use of eigencurves neatly sidesteps the question of map parametrization when performing a BIC analysis \citep[c.f.][]{Kreidberg2018}.
We demonstrate the efficacy of this approach by its application to data for the hot Jupiter HD~189733b and retrieve a map in good agreement with previous results \citep{deWit2012,Majeau2012}.

It is generally better (and can be much faster) to fit data with orthogonal basis functions. In the case of exocartography, the emphasis has so far been to have orthogonal basis \emph{maps}. This is fine for thermal phase curves, for which certain orthogonal basis maps---notably spherical harmonics---produce orthogonal lightcurves---sinusoids of different frequencies.  But eclipses do not share this property: we know of no orthogonal basis maps that produce orthogonal basis lightcurves.  Our eigencurves become sinusoidal for phase-only scenarios, but remain orthogonal even for eclipse mapping.  For planets with small orbital uncertainties, these provide a faster approach to exoplanet mapping.  Moreover, because they are ranked based on their deviation from the uniform planet lightcurve, using the first few eigenmaps ensures that we are maximizing the amount of spatial information we can possibly retrieve from the lightcurves.

While it has been recognized that there are degeneracies between orbital and mapping parameters, we demonstrated an approach to determine whether the uncertainties on system parameters are small enough that they do not limit our mapping ability (as we show is the case for HD~189733b). This can even be done when planning an observation, in order to determine, for example, whether additional radial velocity data are needed to sufficiently reduce uncertainties to the level that a map can be robustly retrieved.

In summary, we recommend that researchers wanting to extract a planet map from their data should:
\begin{enumerate}
\item Produce a set of light curves, corresponding to different thermal emission patterns on the planet. We recommend spherical harmonics and the convenient SPIDERMAN code \citep{Louden2018}.
\item Use principal component analysis to calculate a set of orthogonal ``eigencurves'' (linear combinations of the input light curves) and sort the eigencurves in order of their information content (Section \ref{sec:eigencurves}).
\item Test for the impact of orbital uncertainties on any retrieved map of the planet (Section \ref{sec:orbiterrors}). This can be done by calculating additional sets of light curves, with orbital parameters adjusted by $\pm1 \sigma$, and then running this larger matrix through the PCA. The eigencurves can then be checked to see whether there are any significant differences between the coefficients that multiply input light curves from different orbital realizations and---if there are differences---whether those impact the eigencurves with more/less information content (those that are easier/harder to retrieve from the data).
\item Based on the assessment of orbital uncertainties:
\begin{itemize}
\item If the eigencurves are insensitive to $1\sigma$ differences in orbital realizations, then use the eigencurves to fit the light curve data (Equation \ref{eqn:lightcurve}). Compare the $\chi^2$ values and Bayesian Information Criteria for fits with different numbers and combinations of eigencurves to determine a preferred fit.
\item If the eigencurves are significantly impacted by orbital uncertainties, then they do not form an optimal basis set. In this case it would be sufficient to use something like spherical harmonics in fitting the data. It would still be possible to perform PCA on a representative set of lightcurves drawn from the posterior to determine what spatial information the data were most sensitive to, but only in hindsight.
\end{itemize}
\item Use the calculated eigenmaps, Equation \ref{eqn:finalmap}, and the coefficients from the preferred fit to produce the retrieved planet map. (See Appendices \ref{sec:tbright} and \ref{sec:fstar} for how to translate from relative flux units to brightness temperature, if desired.)
\item In addition to the preferred fit, also report the eigenmaps and any regions unobserved by the data, so that a reader is aware of what spatial information is (un)available in the mapping measurement.
\end{enumerate}

In using our method to retrieve a map of the bright hot Jupiter HD~189733b, we have shown that the orbital elements are known well enough that they do not introduce uncertainties in the planet map. The data are not sufficiently precise to retrieve north-south shifts in the brightness distribution, but we can reliably retrieve the east-west shift and flux amplitude of the dayside hotspot.

Finally, there are a few last warnings that are worth highlighting:
\begin{itemize}
\item The sorted eigenmaps contain the most-retrievable spatial information on the planet, but this does not necessarily bear any resemblance to the actual physical state of the atmosphere. The first eigenmaps are generally large-scale spatial patterns; even if the data are of sufficient quality to map these components, no signal will be detected if the planet has no large-scale features.
\item The method we present here does not inherently prevent fits with non-physical properties. In particular, it may be necessary to add limits in the fit that exclude solutions with negative fluxes or temperatures on the planet \citep{Keating2017}. 
Since we know the eigenmaps in advance of the fit, a check for non-negative planet values could be implemented within the MCMC without too much additional computing time. However, removing the unphysical fits may lead to under-estimated measurement uncertainties, by imposing both physically based and data-driven constraints. Running the MCMC blind to physical constraints may also be informative, in that a large fraction of unphysical solutions in the posteriors may perhaps warn us that the parameterization used is not a good one for the observation. Regardless of the tests used in analysis and model comparison, the final preferred solution must be physically possible and so at some point there must be a rejection check in place to require positive local planet temperatures.
\item In high-precision photometry it is common to decorrelate detector behavior simultaneously with the fit for astrophysical parameters. This means that one's choice of astrophysical model impacts the detrended lightcurve, as recently shown by \citet{Kreidberg2018}. In such a scenario, different eigencurves may be correlated with each other via the parameters of the detector model. While eigencurves are necessarily an orthogonal set, it is still important to beware of the inherent messiness of this type of data analysis in practice.
\end{itemize}

We look forward to the plethora of exoplanet maps that will soon be enabled by JWST. While we have recommended a method for creating horizontal maps at a particular wavelength (or filter band), we expect that many spectral observations with JWST will produce mapping quality data. In order to exploit the full three-dimensional mapping potential of JWST, the path forward will necessarily require a robust way to unite the retrieval of horizontal (time) and vertical (spectral) information within a consistent framework.

\acknowledgments

We thank Eric Agol for sharing the reduced Spitzer eclipse and phase observations of HD~189733b. The first author thanks Erin May for showing her how to use Python. We are also very grateful to Tom Louden for helping us use a pre-release version of SPIDERMAN and being so responsive to requests for help. Support for this work was provided by NASA through an award issued by JPL/Caltech. Finally, we thank the anonymous referee for their thoughtful and constructive comments, which helped to improve this paper.

\software{AASTeX6.1 \citep{aastex61}, emcee \citep{emcee}, iPython \citep{iPython}, Matplotlib \citep{Matplotlib}, NumPy \citep{numpy}, SciPy \citep{SciPy}, SPIDERMAN \citep{spiderman}}

\appendix

\section{Converting from units of flux to brightness temperature} \label{sec:tbright}

The data we analyze in Section~\ref{sec:data} are presented as a flux ratio between the planet at the star which, since they are both effectively the same distance from us, is simply
\begin{equation}
\frac{F_p(t)}{F_*} = \left( \frac{R_p}{R_*}\right)^2 \frac{\oint{V(\theta,\phi,t)M_p(\theta,\phi)}d\Omega}{\oint{V(\theta,\phi)M_*(\theta,\phi)}d\Omega},
\end{equation}
where $M$ is the top-of-the-atmosphere flux, which is weighted by the visibility function $V=\max [\cos \gamma_0,0]$, with $\gamma_0$ being the angle away from the point on the sphere facing the observer \citep{Cowan2013a}. If $\theta$ and $\phi$ are the latitude and longitude, respectively, the solid angle is expressed as $d\Omega=\cos \theta d\theta d\phi$. Since we assume the stellar flux to be spatially uniform and constant in time, we can subsume it into the integral,
\begin{equation}
\frac{F_p(t)}{F_*} = \oint{V(\theta,\phi,t)M_{p,n}(\theta, \phi)d\Omega}, 
\label{eqn:fratio}
\end{equation}
where the units of the planet's flux map are now renormalized to be relative to the stellar flux, with a factor to account for the relative sizes of the planet and star:
\begin{equation}
M_{p,n}(\theta, \phi) = \left( \frac{R_p}{R_*}\right)^2 \left(\frac{M_p(\theta,\phi)}{\pi M_*}\right). \label{eqn:fluxunits}
\end{equation}
A factor of $\pi$ appears in the denominator because $\oint V d\Omega=\pi$. Our retrieved planet map ($Z_p$ in Equation~\ref{eqn:finalmap}) is exactly equivalent to the form of the map we have derived here.

We can then use this normalization to convert from the flux units of our retrieved planet map to a brightness temperature on the planet (at 8 micron), by setting the planet and stellar fluxes to the Planck function, $M_{p}(\theta, \phi) = B_\lambda (T_p(\theta,\phi))$ and $M_* = B_\lambda(T_*)$, where $B_\lambda(T) = (2hc^2/\lambda^5)/(\exp[hc/\lambda k T] -1)$.

Rearranging this equation to solve for the brightness temperature of the planet, we arrive at
\begin{equation}
T_p(\theta,\phi) = (hc/\lambda k) / \ln \left[1+\left(\frac{R_p}{R_s} \right)^2 \frac{\exp[hc/\lambda k T_s]-1}{\pi Z_p(\theta,\phi)}\right], \label{eqn:tbright}
\end{equation}
where we have explicitly replaced $M_{p,n}$ with our retrieved planet map from a fit to the data, $Z_p$. Using $\lambda=8 \mu$m and $T_s=5052$K (see Table \ref{tab:params}), as well as the minor correction described below, we produce the brightness temperature map shown in Figures~\ref{fig:HD189_map} and \ref{fig:lon_profiles}.

\section{Including the stellar flux correction term} \label{sec:fstar}

The data we analyze have been adjusted to remove the stellar flux and to normalize the planet flux relative to stellar. So the data units are:
\begin{equation}
F_d = \frac{(F_p + F_s)-F_c}{F_c} = \frac{F_p}{F_c} + \left[\frac{F_s}{F_c} -1 \right],
\end{equation}
where $F_s$ is the actual stellar flux and $F_c$ is what it was estimated to be. Since our fit (using Equation~\ref{eqn:lightcurve}) has a non-zero value for $F_{*,corr}$, this means that $F_c \ne F_s$, with $F_{*,corr}=\frac{F_s}{F_c} -1$. The value for $F_{*,corr}$ from our preferred fit is $(4.5 \pm 0.4) \times 10^{-4}$, meaning that we only need to impose a very small correction to the stellar flux. This is achieved by multiplying the right hand side of Equation \ref{eqn:fratio} by a correction factor of $(F_{*,corr}+1)$, which then follows through the derivation to end up as a multiplicative factor of $Z_p$ in Equation \ref{eqn:tbright}.

\bibliography{biblio.bib}

\end{document}